# Construction of a Prototype Spark Chamber

Jack Collins
Peterhouse, University of Cambridge
Cavendish Laboratory
12th October 2009

**Abstract**

A small demonstration spark chamber is to be built at the Cavendish laboratory. A prototype chamber consisting of five 20x22.5cm plates has been built and descriptions of its properties and construction are given, while a second chamber with a somewhat novel design is nearly complete. A discussion of the issues surrounding the final design is presented, and recommendations are made in light of the results of this work.

## 1. Introduction

The spark chamber is a track-locating particle detector which saw much use in elementary particle physics research during the 1960s. The device typically consists of a stack of electrically conducting plates separated by around 1cm and immersed in a noble gas, and the trail left by ionizing radiation is indicated by a series of bright sparks following the application of a triggered high voltage pulse. The advantages of this type of detector compared to others of its time were its low cost and ease of construction, its good spatial and time resolutions, and the ease of selecting specific events on which to trigger. Spark chambers have therefore been widely used in particle physics experiments and have been particularly useful in studying the interactions of particles with the chamber electrodes themselves, and in studying rare events[*]. Spark chambers were ultimately superseded by superior drift chambers and solid-state detectors, and having fallen out of favour with particle physicists those few that remain (or have been built since their heyday) have been relegated to demonstration experiments.

Yet, perhaps this is the role they were always destined for. Not only are these chambers relatively cheap and easy to construct, they also make quite an impact on viewers. Their bright, loud sparks are far more immediate than the transient, ghostly tracks left in the small cloud chambers often used in schools, and they can have an arbitrarily large sensitive volume. With this in mind a small band of enthusiastic physicists at the Cavendish laboratory, in collaboration with CHaOS Science Roadshow, aim to build a small cosmic ray spark chamber which can fit in the back of a van (i.e. around $0.5m^3$) for outreach purposes. This chamber will become part of the CHaOS Science Roadshow, a multi-week summer event which tours the country reaching thousands of people, as well as other outreach events in the East of England. It is hoped that using this chamber will help to inspire young people and enhance the public understanding of science, and perhaps to encourage more children to consider a career in physics.

The project is in its early stages and the final design has yet to be decided. It was the purpose of this current work to build a functional prototype in order to test various design ideas, and to test the triggering circuit built by a previous physics student to ascertain its potential for use in the final chamber. However this prototype has already had some success of its own as a demonstration chamber, having been popular in both a physics at work event for GCSE students (figure 1.1) and a Cavendish alumni weekend.

## 2. Operating Principles

The primary factors affecting the performance of a spark chamber are the nature of the high voltage pulse, the gap spacing and the gas composition in the chamber. Before discussing these properties, it is worth describing the basic mechanism of sparking in the chamber and defining some basic terms. A charged particle passing through the chamber will liberate electrons due to collisions with gas atoms – typically, 30 or more ion pairs per centimetre are produced in neon. The passage of the particle is registered by a pair of scintillation counters which are mounted above and below the chamber. When the counters register a coincidence, a high voltage pulse on the order of 10kV/cm is applied to every other plate. The seed electrons then accelerate to the anode and if the impressed field is sufficiently high, they initiate a cascade which results in a spark discharge following the avalanche-streamer mechanism described by Raether and Meek and Craggs[2]. A brief description of the mechanism follows, and figure 2.2 illustrates the process.

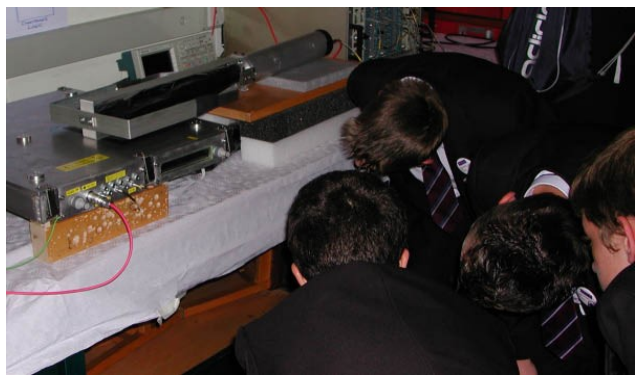

*Figure 1.1 – The spark chamber at 'Physics at work'.*

---

[*] A group at Brookhaven used a 10 ton spark chamber to study rare neutrino interactions in the aluminium plates, relying on the triggering to eliminate unwanted background radiation. Their experiment demonstrated that the electron and muon neutrinos are distinct[1].

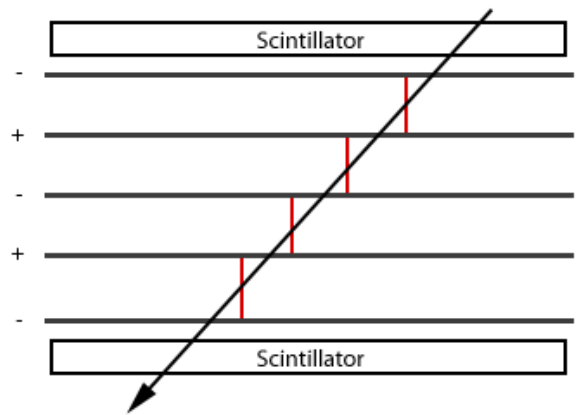

*Figure 2.1 – The path of a cosmic ray is indicated by the arrow. The cosmic ray ionises gas particles in its path, and the resulting ion pairs seed sparks when the high voltage pulse is applied following traversal of the cosmic ray through both scintillation counters.*

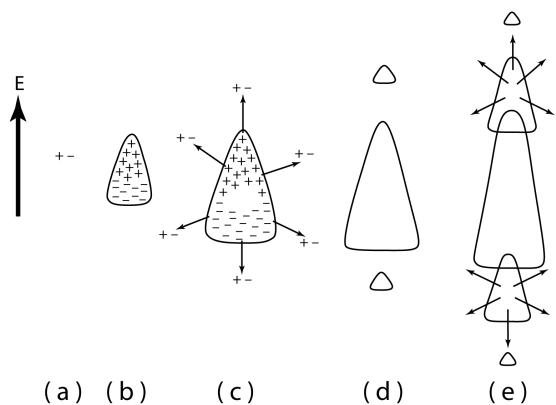

*Figure 2.2 – Streamer formation. (a) Electron liberated by cosmic ray; (b) avalanche formation; (c) recombination results in photon emission producing further photoelectrons; (d) formation of secondary avalanches at the head and tail of the primary; (e) adjacent avalanches merge to form a streamer which develops into a conducting channel[3].*

Collisions between the accelerating electrons and gas atoms result in exponential electron multiplication – each electron produces further electrons at a rate $\alpha$ per centimetre ($\alpha$ being the first Townsend coefficient). This number is a property of the gas and a function of the ratio $E/p$ (where $E$ is the electric field strength and $p$ the pressure), and is in the tens or hundreds under standard spark chamber conditions. This results in an avalanche, with electrons at the head towards the anode, and positive ions at the tail. The process continues until the space-charge field in the avalanche negates the external electric field preventing further growth of the avalanche – this has been found experimentally to occur when around $10^8$ free electrons have accumulated in the head of the avalanche. Recombination with positive ions follows, resulting in isotropic photon emission which creates further ion pairs outside the primary avalanche. The dipole field from the primary avalanche inhibits the formation of further avalanches produced laterally but encourages them in front and behind. Thus, a series of avalanches are produced which join together head-to-tail to form a streamer. This streamer then rapidly advances to the anode and cathode plates forming a conducting plasma channel for the spark. Only a single electron is required to initiate this process. In practice, several seed electrons may generate primary avalanches in a single gap – these will join together into a single channel if they are close enough, but if the cosmic ray comes in at a large angle several sparks may be seen.

The gap efficiency of a spark chamber is the probability that a spark will form close to the ionised trail left by a charged particle when triggered. The maximum permissible time between passage of a particle through the chamber and application of a high voltage pulse to the live plates such that a spark is produced is called the memory time. The dead time is the interval following successful sparking before the chamber can be used again[†]. If the chamber is triggered during this time, reignition of the old track is likely.

### The high voltage pulse

During the brief interval between the cosmic ray passing through the chamber and the application of the high voltage pulse, free electrons between the plates may be lost due to recombination with their parent ions or capture by electronegative atoms. Helium, neon and argon are the most commonly used spark chamber gases, and these have a low affinity for electrons which means they are unlikely to inhibit spark formation. However, electronegative impurities in the gas will reduce the memory time and cause problems if the delay time of the high voltage pulse is sufficiently long. Times of less than 500ns are typical in research grade chambers, but anything less than 1μs is satisfactory when 100% efficiency isn't required[4,5].

---

[†] More precisely, the memory and dead times are given by the period that must elapse before the probability for sparking drops to some specified value, often 1/e.

Clearing fields of tens or hundreds of volts per centimetre are often employed in spark chambers to clear the gaps of residual ionisation, thus reducing spurious sparking and decreasing the dead time which is useful for operation in intense beams. However this also reduces the memory time and places more demanding limits on the delay time and gas purity. Although I built electrical connections for a clearing field into the chamber, it turned out to be unnecessary for this application.

Electrons can also be cleared from the gap by the leading edge of the high voltage pulse. If the pulse rises too slowly, the electrons will drift in the electric field without initiating an avalanche and will be cleared from the gap before the critical voltage is reached. Accumulated wisdom[6] suggests that while a rise time of a few nanoseconds is desirable, anything up to around 0.1μsec is tolerable. The height of the pulse should also be as great as possible without initiating excessive spurious sparking. This will result in a shorter rise time per kilovolt, and will also increase the speed of spark formation. The pulse should be sufficiently long to initiate the discharge, and since sparking times in neon range from 50ns to 300ns for voltages of 15kV/cm and 5kV/cm respectively[8] a pulse length in the hundreds of nanoseconds is appropriate.

**Gas composition**

Not only do noble gases have a low affinity for electrons, they also have high first Townsend coefficients which minimises spark formation times making them ideal spark chamber gases. The values of α for these gases at 10kV/cm and atmospheric pressure are 40/cm, 65/cm and 10/cm respectively[6]. Argon therefore requires much higher voltages for comparable efficiency, but is also very cheap. Neon is the premium option while helium also gives good performance.

Neon produces a strong red spark, while Helium gives a purple one. Helium/neon mixtures are commonly used in varying ratios, and these make use of the same energy transfer process that is utilised in helium-neon lasers to produce the characteristic red spark of neon[7] – only a few percent neon needs to be added to helium to produce this effect. A similar mechanism also accounts for the Penning effect which has also been successfully utilised in spark chambers. Since argon has a relatively low ionisation energy, argon atoms can be ionised by collisions with excited neon or helium atoms which are in a metastable state energetically higher than the ionisation energy of argon. Therefore, adding a small (~1%) quantity of argon to neon or helium can effectively increase the first Townsend coefficient due to the extra electrons made available in the avalanche thus improving the chamber's efficiency. Small quantities of alcohol vapour added to the spark chamber gas has also been known to improve spark chamber performance by absorbing ultraviolet light responsible for much spurious sparking, and by acting as a quenching agent.

Whatever gas is used, spark chambers are normally operated at or near to atmospheric pressure.

**Construction techniques**

The principle goals in the construction of a chamber are to ensure the plate separations are uniform within and between gaps, to keep a tight seal and so maintain gas purity, and to minimise spurious discharges from sharp plate edges within the gas volume. Furthermore, it is desirable that the electrical connections to the high voltage plates be inductance-free to minimise the rise time of the high voltage pulse on the plates. It is also important for the final demonstration chamber to be rugged and compact, relatively cheap, and for there to be good optical clarity of the sparks by eye.

There are two common types of spark chamber construction, which I shall call 'sandwich type' and 'box type'. In the former assembly, the metal plates (typically aluminium) are glued between perspex frames which are polished for visibility, and it is the thickness of these frames which determines the electrode spacing. These chambers are normally operated with a constant flow of fresh gas since the large number of seals introduce numerous opportunities for gas leakage, while the large surface area to volume ratio means that contamination due to outgassing from the walls of the chamber can pose a problem. Edge discharges aren't a problem for such chambers, since the plate edges are buried within the perspex or exposed to air where there is no tendency for sparking. There is a general feeling within the group that the windows in this type of chamber don't offer sufficient optical clarity, and that this would diminish the impact of a demonstration chamber.

The box-type assembly consists of a stack of plates separated by plastic spacers contained within a larger sealed volume. Alternatively, metal spacers can be used between ground plates and between live plates if alternate rectangular plates overhang at right angles to each other. This type of chamber has certain advantages over the sandwich type – in particular, it is easier to make the chamber gas-tight so it can operate for some time on a single fill of gas, and materials of better optical quality than perspex can be used for the windows. On the other hand, one must take care to avoid spurious discharges at the plate edges where the electric field is strongest. One review paper suggests that the plate edges need to be rounded with a minimum radius of curvature of 3mm to eliminate this[9], but perhaps tighter radii are tolerable. Alternatively, it may be possible to coat the plate edges with some kind of electrical insulator such as an insulating paint or tape.

### 3. The Trigger Circuit

A basic schematic of the overall spark chamber circuit is shown in figure 3.1. A NIM crate was used for the discriminator and coincidence logic, though in the future this task will be handled by a printed circuit board. Once the unit registers a coincidence from both scintillation counters, it sends the triggering circuit into action to deliver the high voltage pulse. The delay time associated with these initial stages are considered negligible, possibly tens of nanoseconds at most.

The triggering circuit was built previously by Philip de Grouchy, and a full account can be found in his report[10]. It is based on a design successfully used in a demonstration chamber at NIKHEF in Amsterdam. The circuit contains two fast switching transistors which amplify the input signal of around 3V from the coincidence unit up to 170V within 80ns. A transformer designed and built for a quick rise time amplifies this up to 4kV, which provides the trigger for a spark gap (figure 3.2). The central electrode of a car spark plug acts as the triggering pin, while the outer electrode is grounded. This spark plug is mounted opposite an adjustable anode pin which connects to the high voltage power supply via a 10MΩ resistor. Ordinarily, the anode potential is insufficient to cause electrical breakdown of the air, but when 4kV is applied to the triggering electrode sparking occurs between the central and outer pins of the spark plug, producing UV rays and hence photoelectrons. This initiates a breakdown in the gap between the anode and ground (provided the anode potential is sufficiently high), and it is the resulting spark which triggers the chamber. This shorting of the high voltage line to ground causes a step change in the potential of the 2.2nF capacitors in figure 3.1, and this pulse propagates to the chamber plates where a spark should form along the track of a cosmic ray. In an instant, the circuit goes from being purely capacitive to purely conductive.

The triggering/spark gap circuit was reported to to be capable of raising up to 6.7kV. It appears that this limit was caused by shorting of the 10MΩ resistor via a metal tab on the circuit board. With this removed the circuit is capable of 10kV, and further simple adjustments to the circuit layout are possible to raise this significantly further still. The circuit was also reported to have a delay time of 580±100ns and a repetition rate of ~10Hz. An artificial dead time of 125ms has been built into the coincidence logic to prevent excessively rapid triggering.

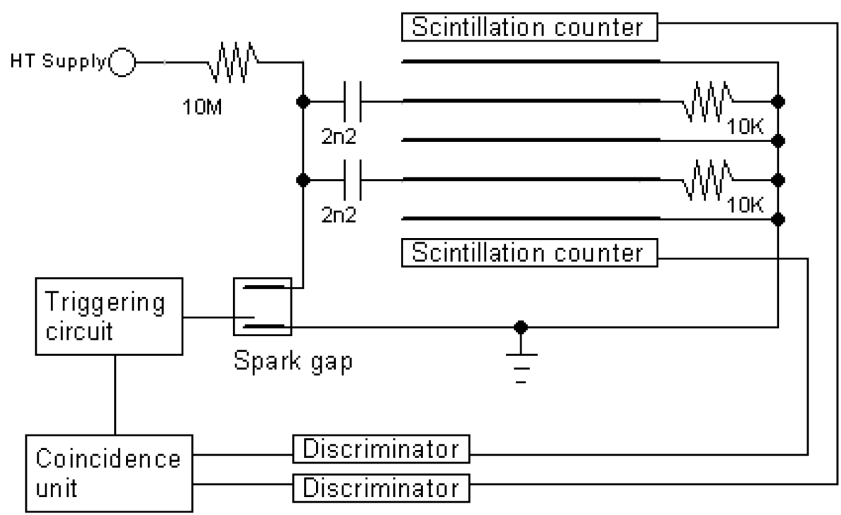

*Figure 3.1 – Spark chamber schematic.*

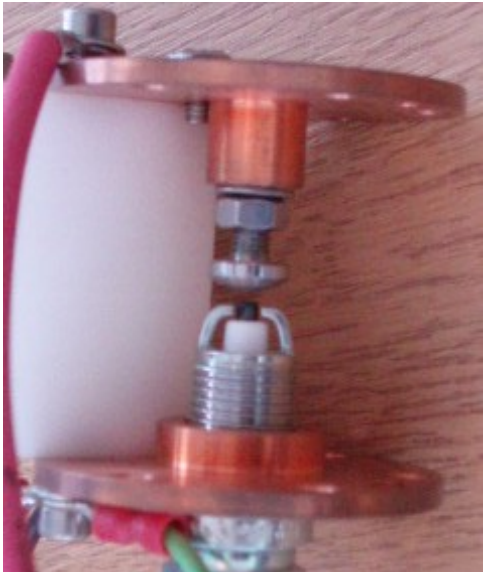

*Figure 3.2 – The spark gap.*

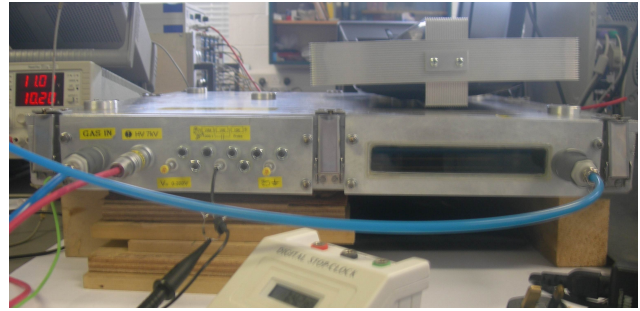

*Figure 4.1 – View of the front of the chamber.*

## 4. Construction of the Mark I Chamber

Since the purpose of this chamber is to act as a test-bed for operating parameters and design ideas, a box type construction was chosen. This approach is versatile since it is possible to alter the stack of plates, for instance to vary gap spacing or the shape of the plates. More importantly it allows us to test the problems posed by spurious discharges at the plate edges, and perhaps to investigate ways of avoiding them should they become an issue.

An old metal box with internal dimensions of 5cm x 50cm x 55cm and walls around 1cm thick was chosen as the case for the chamber. It had previously been used in experiments involving the use of electronics in a gas volume, and having the appropriate electrical connections and a viewing window it was deemed suitable for this work. However the small window (see figure 4.1) means that only a fraction of the box is available for viewing, which makes the setup inefficient as the entire volume must be filled with gas. The top of the box is removable, and is clamped in place with a large O-ring providing the gas seal. The window and electrical connections are on removable plates which are screwed to the front of the box, again with O-rings to provide the gas seal.

Initially the box had a number of very significant gas leaks, which caused a pressure drop of around 1mbar/second when pumped with helium to 150mbar overpressure. The most important of these were around the window, and through small holes at the welded edges. Small blobs of araldite applied strategically on the outside of the box were sufficient to plug these leaks. There were also leaks though the insides of the electrical connections, so all unnecessary connectors were removed and replaced by bolts with nylon washers. Adhesive heat shrink placed over the wire-connector interfaces on the inside of the box provided a good gas seal for the remaining connectors. Finally, liberal application of silicone grease to all O-rings minimised any possible leaks there. After these improvements, the leak rate at 150mbar of helium overpressure was approximately 1mbar every 10 minutes.

Since the maximum voltage which could be provided by the trigger circuit was initially thought to be rather low, an 8mm plate separation was considered to be a suitable compromise between electric field strength and viewability. The height of the box then restricts the stack to five plates, which were cut from 1.15mm thick aluminium sheet with dimensions 200x225mm. Sixteen spacers were machined from 8mm diameter delrin rod to a height of 7.89±0.04mm (where ±0.04mm indicates the range rather than the standard deviation). Twelve of these were drilled centrally with a 4.1mm clearance hole while the remaining four were drilled and tapped for an M4 screw. Eight nylon screws were used to hold the assembly together (four of which were cut short) as shown in figure 4.2. The extra washers underneath serve as legs to raise the assembly. The plates were assembled alternately at right angles so all the live plates overhang 2.5cm one way while the ground plates overhang another way. This was a part of early plans for the electrical connections which later became unnecessary. The electrical connections to the plates are now made by a thin tab machined into the plates, onto which a female crimp connector can be pushed.

Although there are two live plates, there is only one high voltage electrical connection into the box. It is possible to power all the live plates of a spark chamber from a single capacitor, but it is common in such chambers for an early spark to steal much of the capacitor charge and thereby reduce the efficiency for further sparking. The simplest and most effective way to decouple separate live plates is to have a capacitor for each (as in figure 3.1). It is therefore necessary to place the 2.2nF capacitors and 10kΩ resistors for the live plates within the chamber itself, but making this work proved challenging. This is because a circuit which can hold off high voltages in air may spark or leak current by corona discharge in a noble gas. The first circuit can be seen on the left of figure 4.3. Several layers of anti-corona lacquer were sprayed very liberally onto the circuit, in the hope that it would act both as an electrical insulator and as an insulator from the gas. Ultimately, when the circuit was placed in helium gas these measures proved completely inadequate. As charged leaked across the capacitor pins at around 2kV, it became impossible to raise the voltage any further. Figure 4.4 is a picture taken of corona discharge within the chamber in helium.

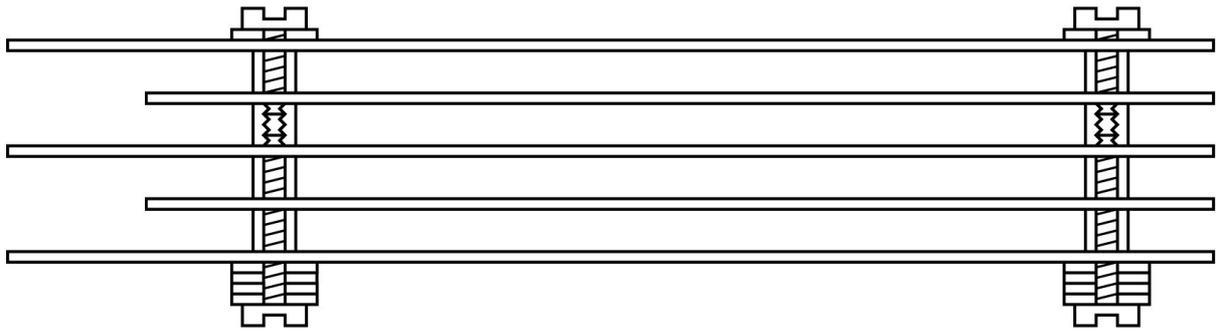

*Figure 4.2 – The plate assembly.*

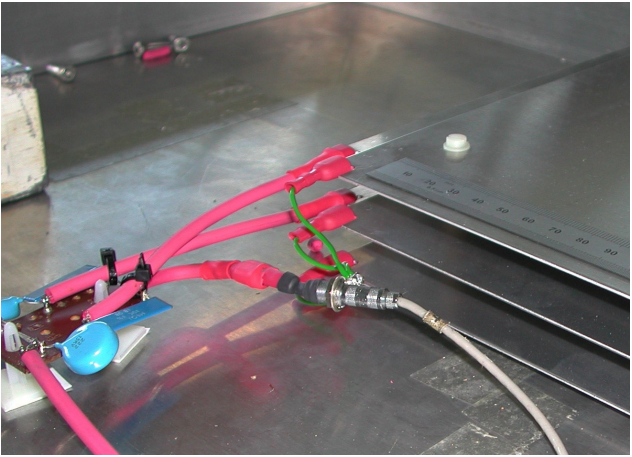

*Figure 4.3 – Electrical connections to the plates.*

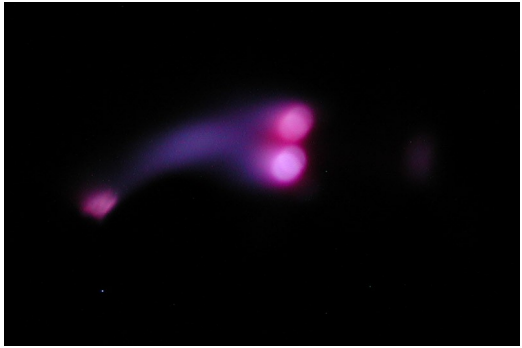

*Figure 4.4 – Corona discharge in the electronics.*

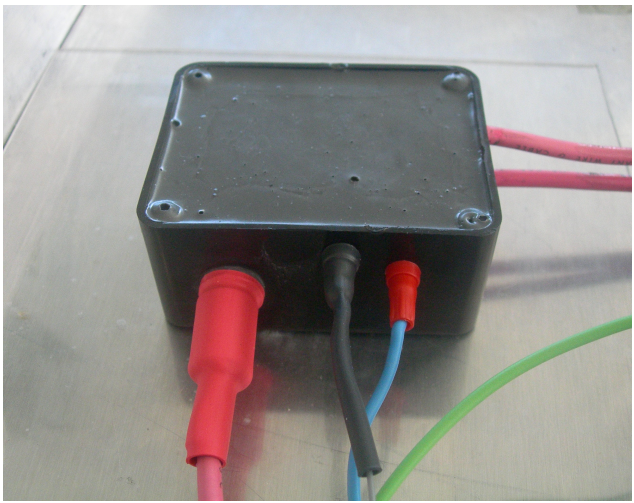

*Figure 4.5 – The potted electronics box.*

It therefore became necessary to isolate the electronics from the gas within the chamber. Some new electronics were assembled and placed in a plastic container (figure 4.5). The container was drilled to accommodate a high voltage connector, two 2mm connectors for the clearing field[*] and probe point[†], and two high voltage leads to the live plates. A polyurethane resin was used to fill the container. This comes as two components which harden after mixing. The hardened resin has a dielectric strength of 16kV/mm, but most importantly it completely isolates the electrical components from the gas – its relatively low viscosity means that few gaps or bubbles are formed. The high voltage connector was taken from an old laser. The female part posed no problem since it was embedded in the resin inside the box, however some work was required to prevent exposure of the soldered joint in the male part to the gas (see figure 4.6). Unfortunately, at 7kV sparking occurs between this connector and the adjacent clearing voltage and probe connections. The extra layer of heat shrink shown in figure 4.5 does little to prevent this. It is not clear whether the sparks travel between the mating surfaces of the male and female high voltage connectors, through the joint between the plastic sleeve and the casing of the male connector, or in some other way. Nonetheless, the 7kV which could be sustained proved to be sufficient to operate the chamber.

---

[*] While a clearing field was never used in this investigation, the capability was built in. The two 10kΩ resistors of figure 3.1 go to the clearance voltage rather than directly to ground as shown in the diagram. The clearance voltage line goes to a LEMO connector in the chamber wall, which was then shorted to ground via a 50Ω resistor for the purposes of this work.
[†] A potential divider was added across one of the 10kΩ resistors. This involved one 10MΩ resistor within the case and two small resistors outside the chamber, allowing the potential across the plates to be probed safely and conveniently.

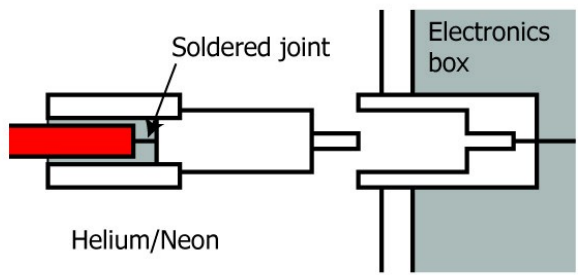

*Figure 4.6 – The high voltage connection to the box. Grey areas represent the potting compound.*

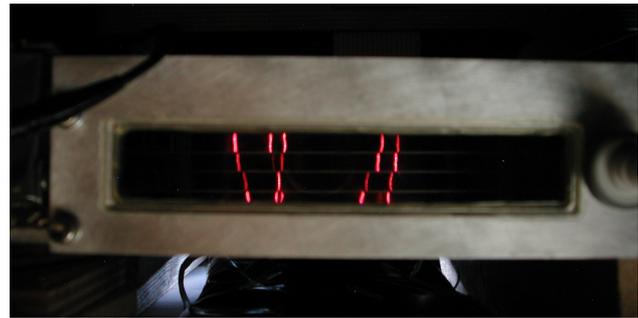

*Figure 4.7 – A long exposure picture. These sparks were not simultaneous, but occurred over several seconds.*

Cylinders of helium and a 70/30% neon helium mixture were available for this work. The gas rig was set up to provide a steady rate of flow through the chamber. Initially it was hoped that the chamber could be flushed of air by evacuation, but early tests ruled this out. The unfortunate shape of the chamber means that evacuation produces an enormous force on the top and bottom, and the substantial flexing of the base of the chamber could eventually result in long term damage. Therefore, the chamber should be flushed with helium and then with the helium/neon mixture. The flow rate can be measured using a pair of flow meters, both of which are calibrated for use with the helium/neon mixture. The exhaust gas is released through a bubbler or a non-return valve to prevent air returning into the chamber.

The chamber was finally ready for use eight weeks into the project, and on the 13$^{th}$ of August it was turned on to produce its first cosmic ray tracks. Figure 4.7 shows one of the early photos taken.

### Observations

During the weeks that elapsed between the plates being assembled and the chamber being made ready for use, the plates were leant on, scratched and generally mishandled – yet this seems to have had little effect on the performance of the chamber. Spurious discharges at the plate edges caused some problems so the corners and edges were rounded and deburred. This made some improvement, though no quantitative analysis was made. However in some places some accidental scratches were made during this process and it is possible that this had an adverse effect in some places. A single layer of kapton tape (dielectric strength 7kV/mm) was stuck around the edges of the bottom two plates – this had no obvious effect, but again no quantitative analysis was made. Curiously, edge sparking rarely occurs where one edge is directly above another but is very common where there are overhanging edges. The edge sparking was a mild nuisance rather than a serious issue (when the gas purity was high and the delay time short), so no further efforts were made in this direction.

Before continuing on to describe my experiments with gas purity and voltage, I will describe some of the general properties of the chamber. Figure 4.9 shows the variations in potential when sparking occurs. The drop in potential on the electrode trace is the high voltage pulse, and the sudden jolt is when sparking occurs. The delay time between coincidence and the beginning of the voltage rise on the plates is 480±50ns, while the rise time is 80ns. The slow rise time is a result of poorly optimised circuitry, some of which was an unavoidable consequence of the overall setup. When the chamber is filled with the helium/neon mixture at around 99.5% purity and operated at 5-6kV, the sparking efficiency is sufficiently good that nearly all events produce clearly recognisable tracks. However, since each live plate powers two gaps, it is common for sparks in one gap to inhibit those in the next, reducing their intensity. The efficiency for multiple tracks is highly variable – some such events are difficult to make out, while others (like that pictured in A1 in the appendix) are clear. It is common for spurious discharges to occur in clumps (A5, A6). This is possibly a consequence of photoelectrons from primary avalanches causing spurious streamers near the plate edges where the electric field is strong, and then these will create yet more photoelectrons in the same region. It is therefore possible that using scintillators a little smaller than the plates themselves would reduce spurious discharges by not triggering on particles which skim the plate edges.

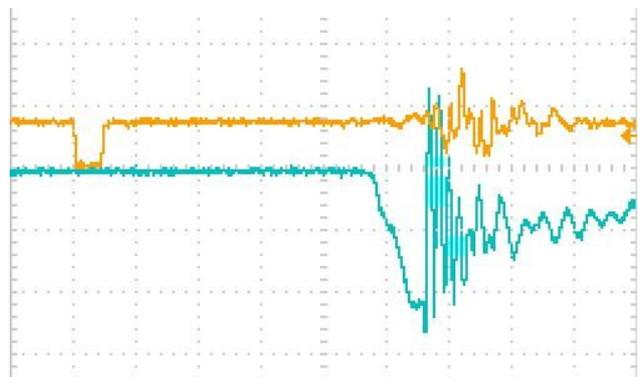

*Figure 4.9 – Spark chamber trace. The orange line is the signal from the coincidence unit, and the blue line is the potential across the chamber plates. The major divisions along the x-axis represent 100ns.*

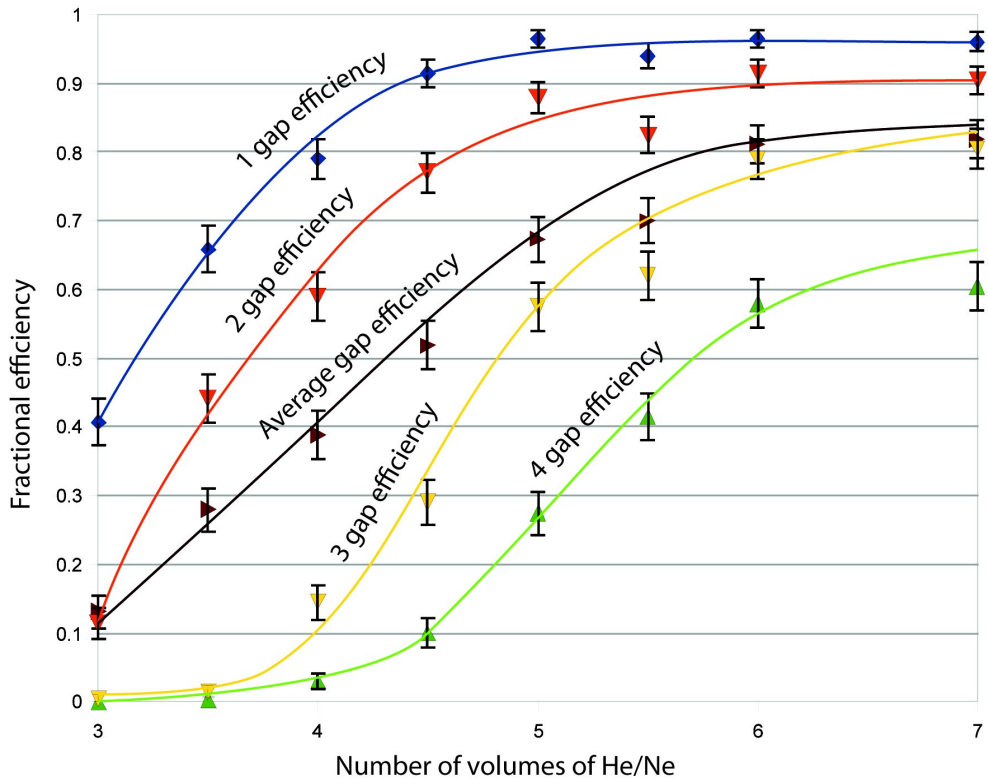

*Figure 4.10 – Spark chamber efficiency. These measurements were taken with a pulse height of 5kV and a delay of 550ns till sparking (this includes the rise time).*

I was curious to see the relationship between gas purity and efficiency. I chose the approach of measuring the n gap efficiency, which I define as the probability of n or more gaps sparking along the trail of ionising radiation. The gas purity was measured by flushing the chamber with the helium/neon gas mixture at a controlled rate. For sufficiently low flow rates, the fraction of the gas in the chamber which remains air after a volume V of noble gas has been flushed through is $e^{-V/U}$, where U is the volume of the chamber*. From now on, I shall call the ratio V/U the 'number of volumes' flushed through the chamber. After 3 volumes have been flushed 5% of the chamber gas is air, while after 7 volumes only 0.09% air remains. The results of this experiment are shown in figure 4.10 above. The average gap efficiency is simply the average probability that any particular gap will spark when triggered. The efficiency measurements were made by watching videos of the spark chamber and tabulating the number of sparks seen per event. It is common, particularly at the lower gas purities, for there to be a

---

* This is a standard exponential relationship. Consider adding ε volumes of gas to the chamber, then removing ε of the mixture. Each time this step is made, the amount of air remaining is multiplied by a factor (1-ε) and so after one volume has been flushed (i.e. 1/ε steps have been made), the fraction remaining in the chamber is

$$\lim_{\epsilon \to 0} (1-\epsilon)^{1/\epsilon} \equiv 1/e$$

|              | 1 gap     | 2 gap     | 3 gap     | 4 gap     | Average gap |
|--------------|-----------|-----------|-----------|-----------|-------------|
| Low voltage  | 0.94±0.02 | 0.93±0.02 | 0.50±0.04 | 0.19±0.03 | 0.64±0.02   |
| Pure helium  | 0.91±0.02 | 0.84±0.03 | 0.49±0.04 | 0.23±0.03 | 0.62±0.02   |

*Figure 4.11 – Spark chamber efficiency in helium and at low voltage.*

bright spark on one side of a live plate and a faint one on the other. Since there is no clear dividing line between the two extremes, deciding which sparks count is subjective. Where there was some uncertainty, it is the number of sparks which would likely be registered by a human eye while watching the chamber live that was recorded, rather than the number of sparks visible in the still picture. Since I didn't make precise measurements of the size and position of the scintillators it is possible that some (or most) of the events which produce no sparks correspond to particles which didn't traverse the sensitive volume of the chamber.

The 3 and 4 gap efficiencies seem to suddenly 'turn on' at around 4-4.5 volumes (99% purity). This can be understood as a consequence of more seed electrons surviving till the application of the high voltage pulse at higher gas purities, which results in less fluctuations in the spark formation time from gap to gap. This also manifests itself in the increased tendency for individual sparks to follow the trajectory of the particle. It seems unusual that the 3 and 4 gap efficiencies should suddenly level off after 6 volumes, and it is worth taking more data at 6.5 and 7.5 volumes to confirm this. This data should be taken as a rough guide to the performance of this chamber, rather than as a definitive study of spark chamber performance in general.

The chamber was also tested with pure helium (7 volumes, with a 5kV pulse and 550ns delay till sparking), and with 7 volumes of the helium/neon mixture at low voltage (3.5kV). The results are tabulated above. It is surprising that the chamber should work at all at such low voltages. According to theory[6], and using measurements of the first Townsend coefficient in neon and helium (Von Engel, 1956) the gap spacing would need to be at least 2cm to spark at that field strength. The chamber works regardless, and with reasonable efficiency. Furthermore, the occurrence of spurious discharges is reduced at low voltages. The efficiency with helium is also reasonable, but somewhat smaller than with the helium/neon mixture.

## 5. Construction of the Mark II Chamber

Given the poor viewing conditions of the prototype chamber, I made plans for the construction of a second which could serve as a temporary demonstration chamber. The 'shelf type' design (figure 5.1) of this chamber can be considered a hybrid of the sandwich and box types. Essentially, the plates are slid into grooves milled in four internal faces of a perspex box. In this way, each gap has a semi-isolated gas volume like the sandwich type chamber, but the actual construction more closely resembles the box type. Gas flow is facilitated by small recesses around the plates machined into alternating ends of the perspex box. The gas is pumped into the bottom, flows through each gap in series, then is released out of the top. Electrical connections are made via screws through the perspex wall which push into the edge of the plates. The machining of the box faces took much longer than anticipated, and so I was unable to personally finish this project. The assembly of the box is kindly being continued by Maurice Goodrick and Rick Shaw at the Cavendish laboratory. This new chamber is nearly ready for action, but ironically edge sparking is proving to be a much greater problem here than it did in the previous chamber.

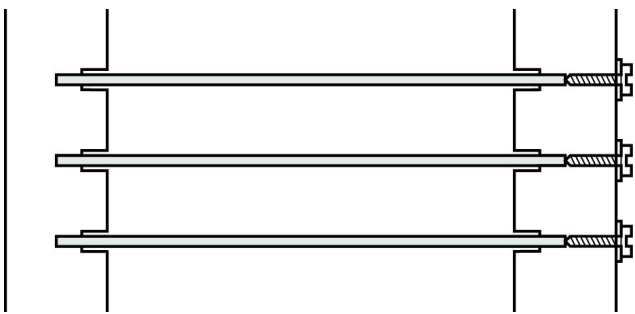

*Figure 5.1 – Shelf type spark chamber. The stepped groove minimises spark tracking along the chamber walls.*

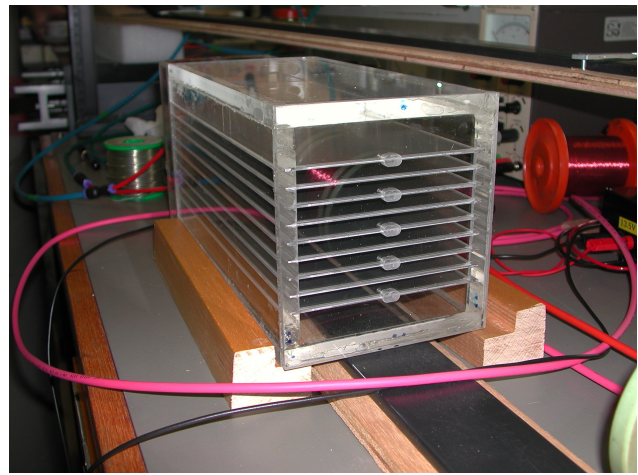

*Figure 5.2 – Shelf type spark chamber.*

## 6. Conclusions and Recommendations

The prototype spark chamber has been a success, both as a testing platform and also in a more limited fashion as a demonstration chamber in its own right. It is hoped that the completion of this phase of the project will hasten progress towards the final goal, and that the contents of this report will be useful in designing the final chamber.

It has been shown that the pulsing circuit built by Philip de Grouchy is capable of triggering a spark chamber successfully. The possibility of buying a commercial trigger still exists, but it is certainly not necessary. While gap uniformity is an important issue, it seems that small scratches on the plate surfaces don't pose a great problem. It would certainly be wise to handle chamber plates carefully, but this indicates that highly ground and polished surfaces are definitely not required. Tolerances on the order of 0.1mm for 1cm gaps seem to be both acceptable and manageable. Edge discharges have been a mild nuisance rather than an incredible problem, and it is possible to reduce this by rounding corners. It may be worth trying thicker plates in future, which can be rounded to a greater radius of curvature. Using larger plates may reduce edge sparking if it is the case that a significant cause is photoelectrons produced by avalanches in the vicinity of the plate edges. If edge sparking problems persist, then insulating materials on the plate edges can be tested using this chamber. It may be worth experimenting with Penning mixtures by the addition of argon to increase the chamber efficiency, and also with alcohol vapour to reduce spurious discharges.

There are many viable designs for the final chamber, but it seems that a simple stack of plates within a larger gas volume could be the most sensible in terms of gas efficiency, simplicity, and visibility. The shelf-type design which I have described may prove to be a viable alternative if the prototype turns out to be successful. In any case, it is recommended that all electrical components are kept outside the gas volume.

## Acknowledgements

I would like to thank everyone at the Cavendish laboratory who helped make this project work. Special thanks to Maurice Goodrick, Saevar Sigurdsson and Rick Shaw for their electrical and machining help and advice, and for their continued efforts in putting together the MKII.